\documentclass[12pt]{article}

\usepackage[letterpaper,margin=1in]{geometry}
\usepackage{amsmath,amssymb}
\usepackage{graphicx}
\usepackage{booktabs}
\usepackage{xurl}
\usepackage[authoryear,round]{natbib}
\usepackage{setspace}
\usepackage{caption}
\usepackage{lineno}

\DeclareGraphicsExtensions{.pdf,.png}
\graphicspath{{./}}
\newcommand{\justifiedpar}{\leftskip=0pt\rightskip=0pt\parfillskip=0pt plus 1fil}
\newcommand{\alttext}[1]{\par{\doublespacing\justifiedpar\noindent Alt text: #1\par}}

\newcommand{\capData}{Benchmark data overview. (a) Ground-truth epicenters of the regional earthquakes and the 2024 $M_w$ 7.4 Hualien aftershocks. The regional earthquakes reach 314~km depth, and the Hualien earthquakes are shallower than 90~km. (b) One 20-minute window of the regional earthquake benchmark noisy set, with the picks of its earthquake and the noise picks of the window plotted against epicentral distance. The clean set has the same window without the noise picks. The noisy set has a median of 178 noise picks per window. (c) Five-minute window in the Hualien benchmark with moved picks, plotted against epicentral distance from the mainshock. Eight earthquakes begin in this window, and the arrivals of three of them overlap.}
\newcommand{\altData}{Panel a is a map of Taiwan with the reference earthquakes of the benchmark as gray dots, the earthquakes of the Hualien set as orange dots along the east coast, and the stations as small open triangles. Panels b and c are scatter plots of distance against time. Panel b shows one 20-minute benchmark window with the picks of one earthquake forming a steep band near 300 seconds and gray noise picks across the window. Panel c shows five minutes of 4 April 2024 with moved picks. Eight triangles above the panel mark the origin times.}

\newcommand{\capFlow}{Consensus association. (a) Steps from the five relocated catalogs to the final catalogs. (b) Vote. Each box is a group of paired events, with one vote per associator. (c) Joined picks of two associators at three stations.}
\newcommand{\altFlow}{Three panel figure. Panel a is a flowchart from five relocated catalogs through pairing, vote counting, joining the picks, merging events that share picks and relocation to duplicate removal and the final catalogs, with a side path from the relocated catalogs straight to duplicate removal. Panel b shows the events of five associators as colored markers along origin time in two shaded groups with five and two votes and a dashed group with one vote, and the consensus events below them. Panel c shows blue and green P and S markers along arrival time at stations 1 to 3, and an arrow to the black consensus picks, which keep the earliest of picks within 0.5 s.}

\newcommand{\capRegional}{False positives and F1 scores on the regional benchmark. (a, b) False positive percentages for (a) the clean set and (b) the noisy set. Each bar shows false positives as a percentage of all scored events, subdivided into split, unmatched, and mislocated events. Numbers at the end of each bar indicate the total percentage. F1 score on the noisy set partitioned by (c) azimuthal gap, (d) depth, and (e) noise picks per window. Numbers above each bin denote ground-truth earthquake counts.}
\newcommand{\altRegional}{Five panel figure. Panels a and b are stacked bar charts of false positive percentages for six association methods on the clean and noisy sets, with black split, gray unmatched and light gray mislocated segments. On clean data, rates range from 3.8 percent for Consensus to 8.3 percent for Backproj3D, with almost all errors being split or mislocated. On noisy data, rates range from 4.2 percent for Consensus to 10.2 percent for REAL, and mislocated events remain the largest type. Panels c to e are line plots of F1 by bins of azimuthal gap, depth, and noise picks. Consensus has the highest F1 in all bins except depths of 120 km or more, where PyOcto is higher.}

\newcommand{\capHualien}{Performance on the Hualien aftershock benchmark (3--8 April 2024). (a, b) Daily F1 scores for (a) recorded picks only and (b) with moved picks added. The number above each day is its count of earthquakes. (c, d) False positive percentage of scored events subdivided into split, unmatched, merged, and mislocated events for (c) recorded picks only and (d) with moved picks added. End of bar values indicate total percentages.}
\newcommand{\altHualien}{Four panel figure. Panels a and b are line plots of F1 for each method on each day from 3 to 8 April 2024, for recorded and moved picks. Consensus has the highest F1 on all days with moved picks, and Backproj3D is 0.004 higher on 3 and 4 April with recorded picks. Panels c and d are stacked bar charts of false positive percentages with black split, gray unmatched, hatched merged and light gray mislocated segments. On recorded picks, rates range from 3.7 percent for Consensus to 7.1 percent for PyOcto. With moved picks added, rates rise from 11.5 percent for Consensus to 46.8 percent for Backproj3D, and unmatched events are most of the false positives of Backproj3D and REAL.}

\begin{document}

\doublespacing

\begin{center}
{\Large\bfseries QuakeConsensus: Reducing false detections in automatic earthquake catalogs through consensus phase association\par}
\bigskip
Wu-Yu Liao$^{1,*}$ and En-Jui Lee$^{2}$
\end{center}

\noindent $^{1}$Institute of Earth Sciences, Academia Sinica, Taiwan

\noindent $^{2}$Department of Earth Sciences, National Cheng Kung University, Taiwan

\noindent ORCID: 0000-0003-2353-7006 (Wu-Yu Liao) and 0000-0003-1545-1640 (En-Jui Lee)

\noindent $^{*}$Corresponding author: Wu-Yu Liao, Institute of Earth Sciences, Academia Sinica, 128, Sec.~2, Academia Road, Nangang, Taipei 11529, Taiwan. Email: tso1257771@gmail.com

\noindent Declaration of Competing Interests: The authors developed Backproj3D and the consensus method compared in this study and declare no other competing interests.

\noindent Word count: 3,264 (abstract, main text, Data and Resources, and Declaration of Competing Interests)


\clearpage

\section{Abstract}

Automatic earthquake catalogs rely on phase associators to group the picks of a phase picker into events. Associators also produce false events that inflate the catalog. We measure the false events of five associators, PyOcto, GaMMA, REAL, GENIE, and backprojection, on two benchmarks built from observed phase picks in Taiwan. The first benchmark isolates 28,129 cataloged earthquakes in 20-minute windows, adding noise by shifting unassociated picks independently. The second evaluates 11,917 earthquakes from the 2024 $M_w$ 7.4 Hualien aftershock sequence, shifting unassociated P and S pairs together. We also propose a consensus association strategy that keeps an event found by at least two associators. With noise picks, single associators produce 5.3~\% to 10.2~\% false positives on the regional benchmark and 11.8~\% to 46.8~\% on the Hualien benchmark, whereas the consensus catalog reduces these false positive percentages to 4.2~\% and 11.5~\%, respectively. The consensus has the highest F1 score and the smallest percentage of false positives on both benchmarks.

\section{Introduction}

Deep learning phase pickers now serve as the common tool for detecting P and S arrivals \citep{ZhuBeroza2019PhaseNet, mousavi2020eqtransformer, liao2021ARRU, Liao2022RedPan}, and catalogs built from their picks list many more earthquakes than the catalogs built from manual readings in the same regions \citep{Sun2024Chihshang, Yang2025Hualien, Liao2025Dapu, Lai2025Dapu, Lee2020GAMMA, Liao2022Backproj}. A phase associator decides which picks belong to the same earthquake and estimates a first hypocenter. REAL counts the picks that fit trial origins on a grid \citep{Zhang2019REAL}, GaMMA fits a Bayesian Gaussian mixture model to the picks \citep{Zhu2022GaMMA}, PyOcto searches the space of origins with an octree \citep{Munchmeyer2024PyOcto}, GENIE uses a graph neural network trained on synthetic arrivals \citep{McBrearty2023GENIE}, and backprojection stacks pick probabilities onto a three dimensional source grid \citep{Lee2020GAMMA, Liao2022Backproj, Liao2025Dapu}.

An associator can split the picks of one earthquake into multiple events, merge the picks of two earthquakes into one event, build an event from picks that belong to no earthquake, or place a real earthquake far from its true hypocenter. In an automatic catalog, every such event enters the seismicity rate, the magnitude calculation \citep{Liao2026TWmag}, and the completeness magnitude \citep{GutenbergRichter1944, WiemerWyss2000}. Combining the catalogs of two associators to recover missed earthquakes also combines their false events.

Developers usually evaluate an associator on synthetic arrivals or on a recorded sequence compared with an existing catalog \citep{Zhang2019REAL, Zhu2022GaMMA, McBrearty2023GENIE, Munchmeyer2024PyOcto}. Synthetic arrivals lack the noise picks of a real picker, and an event missing from a catalog may be a small earthquake that analysts did not locate or a false event. A test of false events therefore needs real picks in which the earthquake of every pick is known.

Here we build two such benchmarks from the picks of the RED-PAN picker \citep{Liao2022RedPan} in Taiwan and count the false events of five associators by type. The first samples the regional seismicity, and the second covers the first six days of the 2024 $M_w$ 7.4 Hualien aftershock sequence, when earthquakes might overlap in time and space. Both are tested with the earthquake picks alone and with unrelated picks added. We also test a consensus catalog that keeps an event only when at least two associators find it.

\section{Data and Methods}

\subsection{Benchmark Datasets}

Both benchmarks use the RED-PAN P and S picks on continuous records from 2012 to 2025 at 752 station sites with 1,387 instruments (see Data and Resources). Most stations belong to the Taiwan Strong Motion Instrumentation Program (TSMIP, \citealp{liu1999TSMIP}), the Central Weather Administration Seismographic Network (CWASN, \citealp{CWASN}) and the Broadband Array in Taiwan for Seismology (BATS, \citealp{BATS1994}). The reference catalog is the Central Weather Administration (CWA) catalog from the Geophysical Database Management System (GDMS, \citealp{Shin2013GDMS}; see Data and Resources). Travel times come from the 3D velocity model of \citet{Lee2025VelModel} (see Data and Resources) with a time correction for each station site.

\subsubsection{Regional Seismicity Benchmark}

For each cataloged earthquake, we predict the P and S arrival times at the stations within a distance set by its farthest manual pick in the CWA archive, which is not public. These prioritized manual picks and the RED-PAN pick nearest to each prediction, all within 2.0~s for P and 4.0~s for S of the prediction, form the pick set of the earthquake. Of the earthquakes with at least six picks on four or more stations, we keep the one with the most picks in each occupied cell of 0.05° in latitude and longitude and 5~km in depth (Figure~\ref{fig:data}a). We relocate each with ADLoc \citep{Zhu2025ADLoc} to 3D travel time tables, which rejects arrivals with residuals over 1.5~s, and score the 28,129 earthquakes whose solution keeps at least six picks and lies within 3~s and 30~km of the catalog solution.

Each earthquake is placed alone in a window of 1,200~s, 300~s after the window start, with all its picks shifted by the same time. These windows form the clean set, in which an associator can split or mislocate an earthquake but cannot build an event from unrelated picks.

The noisy set adds noise picks to each window (Figure~\ref{fig:data}b). These are the RED-PAN picks recorded on the date of the earthquake, from 300~s before to 900~s after its origin time, that were assigned to no cataloged earthquake. Picks from 5~s before the first to 20~s after the last pick of any other cataloged earthquake are removed. Each noise pick is shifted by its own random time, uniform between $-300$ and 300~s and wrapped inside the window, which breaks the timing between the P and S picks of each detection.

\subsubsection{Hualien Aftershock Benchmark}

The Hualien benchmark covers 3 to 8 April 2024 and has 11,917 earthquakes. Of these, 2,494 are cataloged by the CWA and have manual and RED-PAN picks, built as in the regional benchmark. The other 9,423 are found by matching the daily picks with the arrival patterns of the cataloged earthquakes of 3 to 30 April. A match needs at least 12 of the 40 template arrivals at the nearest stations, on four or more stations, and half of the template arrivals out to the farthest matched station, within 0.5~s for P and 1.0~s for S, and no cataloged earthquake within 3~s and 30~km. Each match is relocated and kept when its solution lies within 3~s and 30~km of the template. Shifting the picks of each station by an independent random time gives no match under these criteria.

\begin{figure}[!htbp]
\centering
\includegraphics[width=\textwidth,keepaspectratio]{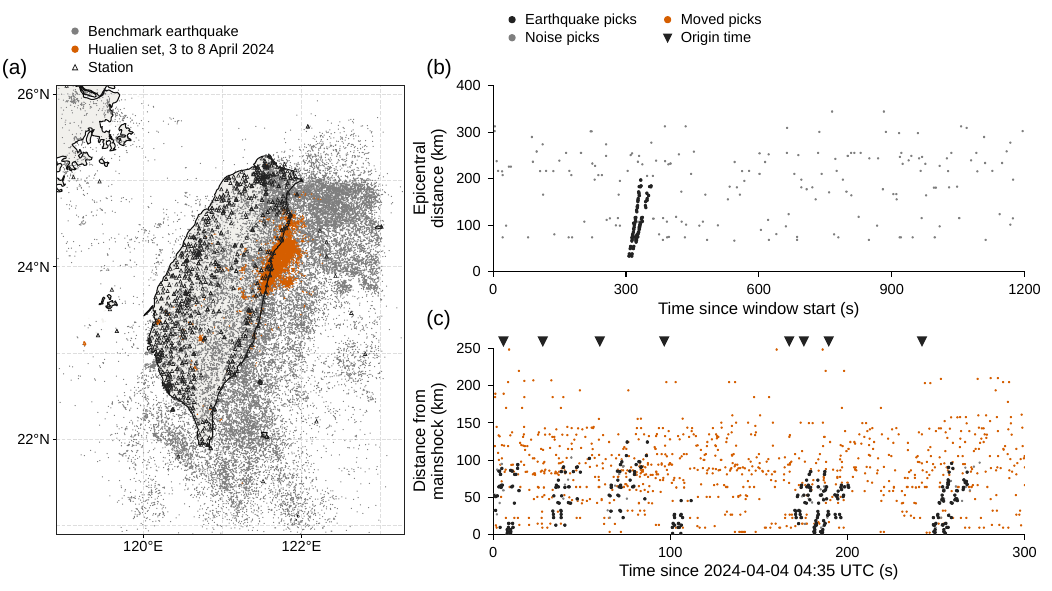}
\caption{\capData}
\alttext{\altData}
\label{fig:data}
\end{figure}

On 3 April the ground truth has 2,979 earthquakes, about 41 in 20~min, which fall to 1,178 by 8 April. The first form of this benchmark contains the recorded picks of the cataloged and matched earthquakes. Picks rejected by relocation stay as unlabeled picks, and earthquakes whose relocation fails keep their picks but are not scored. The second form adds the other RED-PAN detections of the day as moved picks, with paired P and S picks of each detection shifted together by a random time between $-300$ and 300~s, wrapped within the day (Figure~\ref{fig:data}c).

\subsection{Consensus Phase Association}
We evaluate all five associators on the identical set of phase picks. PyOcto \citep{Munchmeyer2024PyOcto}, GaMMA \citep{Zhu2022GaMMA}, and REAL \citep{Zhang2019REAL} use a 1D velocity model derived from the 3D velocity model. We trained GENIE \citep{McBrearty2023GENIE} on synthetic picks from random sources placed according to past seismicity, with P and S times from a travel time network fitted to the 3D velocity model. Our backprojection implementation, Backproj3D \citep{Lee2020GAMMA, Liao2022Backproj, Liao2025Dapu}, stacks pick probabilities from the 20 nearest stations onto an irregular grid with nodes 3~km apart beneath Taiwan.
  
\begin{figure}[!htbp]
\centering
\includegraphics[width=\textwidth,keepaspectratio]{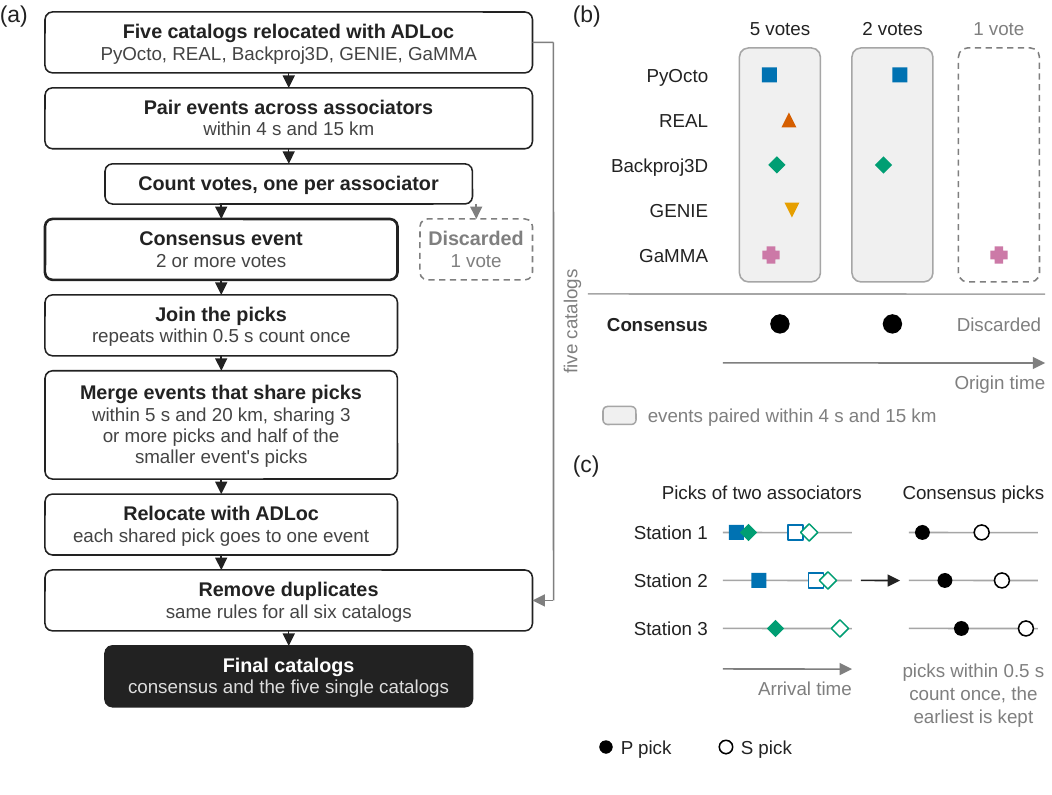}
\caption{\capFlow}
\alttext{\altFlow}
\label{fig:flow}
\end{figure}

We relocate every event of each catalog with ADLoc (Figure~\ref{fig:flow}). Events from different associators are paired if their origin times differ by at most 4~s and their epicenters by at most 15~km. Paired events form a group, and a group with at least 2 of the 5 possible votes becomes a consensus event.

A consensus event takes the union of the picks of its members, in which picks at one station with the same phase and at most 0.5~s apart count once and the earliest is kept. Consensus events that share at least 3 picks and at least half the picks of the smaller event are merged if they lie within 5~s and 20~km in epicentral distance. Each consensus event is then relocated with ADLoc, and a pick shared by two events stays with the event that has both P and S at that station, or else with the event of smaller travel time residual. Events with fewer than 6 picks are dropped before merging and after the shared picks are assigned, and events whose picks changed are relocated again.

Duplicate removal is then applied to the consensus and to the five relocated catalogs, so that all six catalogs are scored with the same rules. It has four steps. First, an event is taken as a split of a larger event and removed when it lies within 15~s and 30~km in epicentral distance of an event with more picks, has at most 70~\% as many picks, and has at least half of its picks within 1.5~s for P or 2.5~s for S of the arrival times predicted from the larger event. The smaller event is kept when most stations shared by the two events have picks more than 1~s apart. Second, of events that lie within 2~s and 25~km in hypocentral distance of each other, only the event with the most picks is kept. Third, a pick shared by two events within 120~s goes to the event with the smaller travel time residual, and events left with fewer than 6 picks are removed. Fourth, each event keeps at most one P and one S pick per station.

\subsection{Evaluation Metrics}
Benchmark phase picks are either assigned to a ground truth earthquake or left unlabeled. An event claims an earthquake if it contains at least 4 picks comprising at least one quarter of the arrivals of that earthquake, or at least 6 picks of that earthquake that constitute at least half of the event. An event is classified as merged if it claims two or more earthquakes, or if it claims one earthquake but derives less than half of its total picks from any single earthquake. Among the other events that derive at least half of their picks from one earthquake, the event with the most picks is designated as the primary event if it claims that earthquake, while the remainder are classified as split events. The remaining events are unmatched when at least half of their picks are unlabeled, and merged otherwise. The primary events of earthquakes outside the ground truth are not scored. A primary event within 2~s and 20~km of the ground truth is a true positive. On the regional benchmark the 20~km limit is about ten times the median location error of the true positives. All other scored events, the split, merged, unmatched, and mislocated primary events, are false positives. A ground-truth earthquake without a true positive is a false negative. Precision $P$, recall $R$, and F1 score are defined as
\begin{equation}
P = N_{\mathrm{TP}}/(N_{\mathrm{TP}}+N_{\mathrm{FP}}), \qquad R = N_{\mathrm{TP}}/N_{\mathrm{EQ}}, \qquad \mathrm{F1} = 2PR/(P+R),
\end{equation}
where $N_{\mathrm{TP}}$ and $N_{\mathrm{FP}}$ are the numbers of true and false positives in a catalog, and $N_{\mathrm{EQ}}$ is the number of ground-truth earthquakes. The sum $N_{\mathrm{TP}}+N_{\mathrm{FP}}$ is the number of scored events of a catalog, and $1-P$ is its false positive proportion. For the regional benchmark, we also score each catalog in bins of the azimuthal gap and depth of the ground-truth earthquakes and of the noise picks per window.

\section{Results}

\subsection{Regional Seismicity Benchmark}

The consensus has the highest precision, recall and F1 on the clean set, and PyOcto has the highest F1 of the single associators (Table~\ref{tab:scores}). Adding noise picks lowers the F1 by 0.007 for the consensus and at most 0.023 for the single associators.

\begin{figure}[!htbp]
\centering
\includegraphics[width=\textwidth,keepaspectratio]{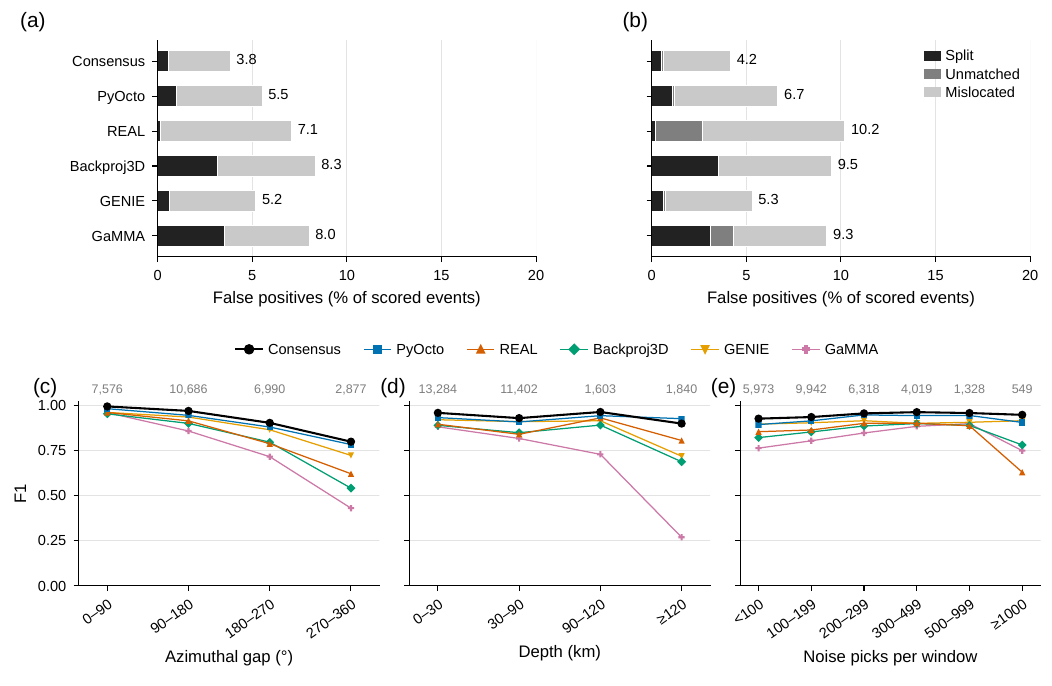}
\caption{\capRegional}
\alttext{\altRegional}
\label{fig:regional}
\end{figure}

Without noise picks, false positives make up 3.8~\% of the consensus events and 5.2~\% to 8.3~\% of the events of the single associators (Figure~\ref{fig:regional}a). Nearly all are split or mislocated earthquakes, because these windows contain no unrelated picks. With noise picks, the percentages rise to 4.2~\% for the consensus and 5.3~\% to 10.2~\% for the single associators (Figure~\ref{fig:regional}b). REAL builds 653 unmatched events from the noise picks and the consensus 20, and mislocated earthquakes remain most of the false positives of the consensus.

\begin{table}[!htbp]
\caption{Scores of the six catalogs on the two benchmarks.}
\label{tab:scores}
\centering
\setlength{\tabcolsep}{4pt}
\begin{tabular}{llrrrrrrrr}
\toprule
& & \multicolumn{4}{c}{Earthquake picks only} & \multicolumn{4}{c}{With noise picks added} \\
\cmidrule(lr){3-6} \cmidrule(lr){7-10}
Benchmark & Method & Events & P & R & F1 & Events & P & R & F1 \\
\midrule
Regional & Consensus & 27,380 & 0.962 & 0.936 & 0.949 & 27,201 & 0.958 & 0.927 & 0.942 \\
& PyOcto & 27,580 & 0.945 & 0.926 & 0.936 & 27,447 & 0.933 & 0.911 & 0.922 \\
& REAL & 25,985 & 0.929 & 0.858 & 0.892 & 26,354 & 0.898 & 0.841 & 0.869 \\
& Backproj3D & 25,636 & 0.917 & 0.836 & 0.874 & 25,468 & 0.905 & 0.819 & 0.860 \\
& GENIE & 25,738 & 0.948 & 0.868 & 0.906 & 25,665 & 0.947 & 0.864 & 0.904 \\
& GaMMA & 22,636 & 0.920 & 0.740 & 0.820 & 23,200 & 0.907 & 0.748 & 0.820 \\
\addlinespace
Hualien & Consensus & 11,263 & 0.963 & 0.910 & 0.935 & 11,314 & 0.885 & 0.841 & 0.862 \\
& PyOcto & 10,881 & 0.929 & 0.848 & 0.887 & 11,766 & 0.786 & 0.776 & 0.781 \\
& REAL & 9,552 & 0.951 & 0.762 & 0.846 & 13,209 & 0.620 & 0.688 & 0.652 \\
& Backproj3D & 11,230 & 0.960 & 0.905 & 0.931 & 19,910 & 0.532 & 0.889 & 0.666 \\
& GENIE & 9,757 & 0.945 & 0.774 & 0.851 & 8,687 & 0.882 & 0.643 & 0.744 \\
& GaMMA & 10,494 & 0.943 & 0.831 & 0.883 & 8,412 & 0.557 & 0.393 & 0.461 \\
\bottomrule
\end{tabular}

\medskip
{\doublespacing\justifiedpar\noindent Regional: regional seismicity benchmark, 28,129 earthquakes. Hualien: Hualien aftershock benchmark, 3 to 8 April 2024, 11,917 earthquakes. Earthquake picks only: the clean set of the regional benchmark and the recorded picks of the Hualien benchmark. With noise picks added: the noisy set and the Hualien benchmark with moved picks. Events: scored events, true and false positives. P for precision, R for recall, and F1 for the F1 score, $2PR/(P+R)$.\par}
\end{table}

On the noisy set, the consensus has the highest F1 in every bin of azimuthal gap, depth and noise picks per window, except at depths of 120~km or more, where PyOcto is higher (Figure~\ref{fig:regional}c to e). The false positive percentage of the consensus falls as the number of votes rises, from 27~\% of the events with two votes to 1~\% of those with five. Raising the vote threshold to 3, 4 or 5 would lower the F1 of the consensus to 0.928, 0.896 or 0.802, because recall falls more than precision rises.

\subsection{Hualien Aftershock Benchmark}

\begin{figure}[!htbp]
\centering
\includegraphics[width=\textwidth,keepaspectratio]{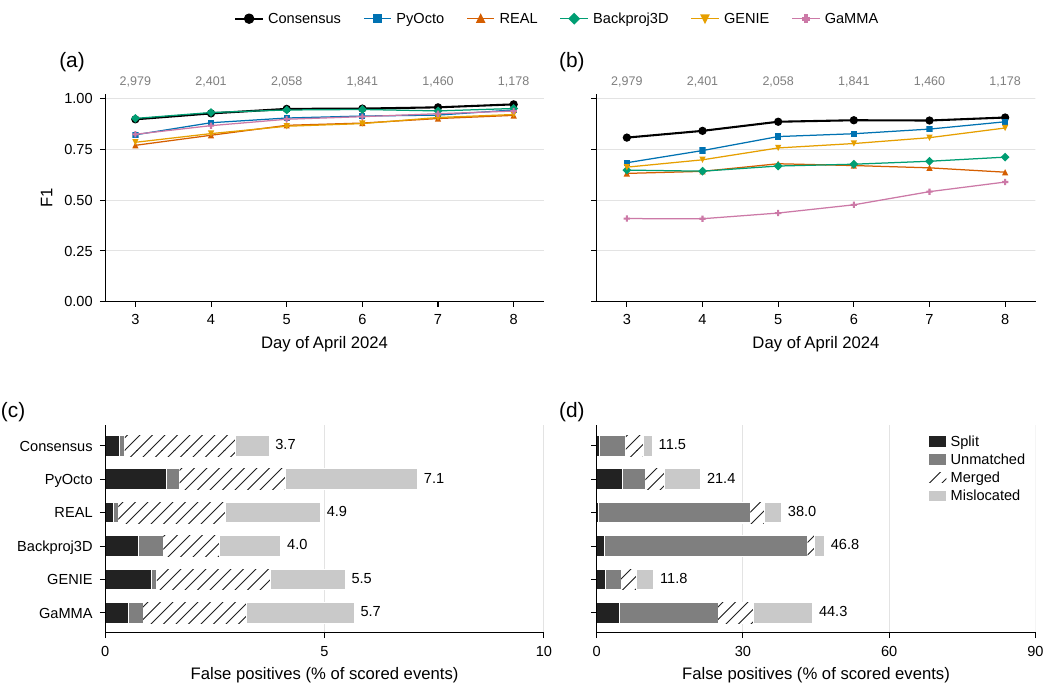}
\caption{\capHualien}
\alttext{\altHualien}
\label{fig:hualien}
\end{figure}

With the recorded picks only, the consensus and Backproj3D have the highest F1 over the six days (Table~\ref{tab:scores}), and Backproj3D is higher by 0.004 on 3 and 4 April, the two days with the most earthquakes (Figure~\ref{fig:hualien}a). Merged and mislocated events are the two largest types of false positive in every catalog (Figure~\ref{fig:hualien}c). With the moved picks, the F1 of the consensus falls by 0.073 and that of the single associators by 0.106 to 0.422, against at most 0.023 for the single associators when noise picks are added on the regional benchmark. The consensus keeps the highest F1 on every day (Figure~\ref{fig:hualien}b), and its false positives rise from 3.7~\% to 11.5~\% of its events. Backproj3D and REAL produce 1.7 and 1.1 events per earthquake, and most of their false positives are unmatched events (Figure~\ref{fig:hualien}d).

\section{Discussion}

\subsection{Comparison of the Associators}

The clean set tests whether an associator keeps the picks of one earthquake together and locates it, the noisy set whether it rejects unrelated picks, and the recorded Hualien picks whether it separates overlapping earthquakes. Split, mislocated, unmatched and merged events mark the failure of each ability in turn. Since all catalogs are relocated with ADLoc, differences in mislocated events come from the picks each associator assigns.

No single associator is best in all three tests. PyOcto leads the single associators on the regional benchmark but has the most false positives on the recorded Hualien picks (Figure~\ref{fig:hualien}c). Backproj3D merges the fewest aftershocks and leads on the recorded Hualien picks, but has low recall on the regional benchmark. REAL splits the fewest earthquakes but has the largest percentage of mislocated events on the clean set (Figure~\ref{fig:regional}a) and the lowest recall on the recorded picks. GENIE has the fewest false positives of the single associators except on the recorded Hualien picks, but a lower recall than PyOcto (Figures~\ref{fig:regional}a, b and \ref{fig:hualien}d). GaMMA has the lowest recall on the clean set and loses the most F1 with the moved picks (Table~\ref{tab:scores}). With the moved picks, unmatched events make up most false positives of Backproj3D and REAL (Figure~\ref{fig:hualien}d), consistent with both grid searches accepting unrelated P and S pairs.

\subsection{Effect of Consensus Voting}

The consensus keeps most of these strengths. It has the highest precision and F1 in all four tests, and the highest recall in all but the test with moved picks, where about half of the Backproj3D events are false positives (Table~\ref{tab:scores}). This is consistent with the associators failing on different events. An unmatched event of one associator rarely has a partner within 4~s and 15~km in another catalog, so it receives one vote and is discarded. The vote cannot remove errors that two associators share. Mislocated earthquakes make up most of the consensus false positives on the regional benchmark (Figure~\ref{fig:regional}a, b), and aftershocks merged by several associators stay merged, which is consistent with Backproj3D scoring above the consensus on the two days with the most earthquakes (Figure~\ref{fig:hualien}a, c).

\subsection{Limitations of This Study}

The regional benchmark keeps the best recorded earthquake of each cell, so its earthquakes are larger and better recorded. Most of the Hualien ground truth comes from template matching, which finds only earthquakes with arrival patterns similar to those of cataloged earthquakes. We applied each associator with the settings of our routine catalog or its published defaults, not with settings tuned on the benchmarks. Both benchmarks come from one network and one picker, so the ranking of the associators may differ elsewhere. The vote keeps false positives that two or more associators build. Picks of coherent noise can meet the criteria of every associator, so these false positives depend on the picker. Mislocated and merged false positives contain real earthquakes and need better locations or separated picks, not removal. Event attributes such as the share of nearby stations with a pick may flag split and unmatched events, but a suitable threshold may vary widely with conditions such as station geometry and wave propagation.

\section{Conclusions}

In this study, we evaluated five phase associators and their consensus on two benchmarks built from recorded phase picks in Taiwan. The regional benchmark tests isolated earthquakes with and without noise picks, and the Hualien benchmark tests overlapping aftershocks with recorded and moved picks. Each associator shows a distinct weakness. PyOcto, the best single associator on the regional benchmark, has the most false positives on the recorded aftershock picks, where Backproj3D is best. GENIE misses more earthquakes than PyOcto, REAL has the lowest recall on the recorded aftershock picks, and GaMMA loses the most F1 with the moved picks. The consensus, which keeps an event found by at least two associators, combines most of their strengths and has the highest precision and F1 in all four tests. However, the vote cannot remove errors that two associators share, mainly mislocated earthquakes and merged aftershocks. These results indicate that agreement among existing associators reduces false positives without tuning any of them.

\section{Data and Resources}

The CWA catalog is available from GDMS (\url{gdms.cwa.gov.tw}), which also provided the CWA waveforms and station inventory. We cannot redistribute the waveforms, the station inventory, or the picks and event tables derived from them. The velocity model of \citet{Lee2025VelModel}, \texttt{LEE2025VpVs}, is available from the Taiwan Earthquake Research Data Center (TECDC, \url{tecdc.earth.sinica.edu.tw/TWtomo/index.php}). RED-PAN, PyOcto, GaMMA, REAL, GENIE and ADLoc are on GitHub (\url{github.com}) as \url{tso1257771/RED-PAN} and its port \url{tso1257771/RED-PAN-Motion}, \url{yetinam/pyocto}, \url{AI4EPS/GaMMA}, \url{Dal-mzhang/REAL}, \url{imcbrearty/GENIE} and \url{AI4EPS/ADLoc}. The Backproj3D code is not publicly available. Websites were last accessed in October and November 2026. The code will be released at \url{github.com/tso1257771/QuakeConsensus} and deposited with a persistent DOI upon publication.

Claude Code 2.1 with the Claude Opus 5.5 model (Anthropic) and OpenAI Codex CLI 0.169 (OpenAI) assisted with code review, number checks, figure scripts, and language editing, but created or edited no figure image directly. The authors reviewed all changes and take responsibility for the content.

\section{Author Contributions}

Wu-Yu Liao contributed to conceptualization, methodology, software, validation, formal analysis, investigation, data curation, visualization, original draft preparation, review, and editing. En-Jui Lee contributed to supervision, funding acquisition, review, and editing.

\section{Declaration of Competing Interests}

The authors developed Backproj3D and the consensus method compared in this study and declare no other competing interests.

\section{Acknowledgments}

The authors thank the Central Weather Administration for access to GDMS data and station metadata.

En-Jui Lee and Wu-Yu Liao are supported by the National Science and Technology Council (NSTC), Taiwan, under Grant No. 113-2116-M-006-004-MY3. Wu-Yu Liao is supported by the Academia Sinica Postdoctoral Scholar Program (AS-PD-1152-M07-2).

\bibliography{refs}

\section{Mailing Address of the Corresponding Author}

\noindent Wu-Yu Liao\\
Institute of Earth Sciences, Academia Sinica\\
128, Sec.~2, Academia Road, Nangang, Taipei 11529, Taiwan\\
Email: tso1257771@gmail.com

\section{List of Figure Captions}

\noindent Figure~\ref{fig:data}. \capData\par

\noindent Figure~\ref{fig:flow}. \capFlow\par

\noindent Figure~\ref{fig:regional}. \capRegional\par

\noindent Figure~\ref{fig:hualien}. \capHualien\par

\end{document}